\documentclass[runningheads]{llncs}
\usepackage[T1]{fontenc}
\usepackage{graphicx,verbatim}
\usepackage{cite}

\usepackage{amsmath,amssymb,amsfonts}
\usepackage{algorithmic}
\usepackage{hyperref}
\usepackage{textcomp}
\usepackage{url} 

\begin{document}
\title{Decoding the Alzheimer's Continuum: Interpretable Multi-Gate Routing for Diagnosis and Transition Prediction}
\titlerunning{Decoding the Alzheimer's Continuum}
%

\author{Yufeng Jiang\textsuperscript{*}\inst{1} \and
Hexiao Ding\textsuperscript{*}\inst{1} \and
Hongzhao Chen\textsuperscript{*}\inst{1} \and
Jing Lan\inst{1} \and
Xinzhi Teng\inst{1} \and
Gerald W.Y. Cheng\inst{1} \and
Yunlin Mao\inst{1} \and
Zongxi Li\inst{2} \and
Haoran Xie\inst{2} \and
Jung Sun Yoo\textsuperscript{\dag}\inst{1} \and
Jing Cai\textsuperscript{\dag}\inst{1}}

\authorrunning{Jiang, Ding, Chen et al.}

\institute{Department of Health Technology and Informatics, \\
Hong Kong Polytechnic University, Hong Kong SAR, China \\
\email{\{yufeng.jiang, hexiao.ding, hongzhao.chen, jing-hti.lan, yunlin.mao\}@connect.polyu.hk} \\
\email{\{xinzhi.x.teng, wai-yeung.cheng, jungsun.yoo, jing.cai\}@polyu.edu.hk}
\and
Division of Artificial Intelligence, School of Data Science, \\ Lingnan University, Hong Kong SAR, China \\
\email{\{zongxili, hrxie\}@ln.edu.hk} \\
\smallskip
\textsuperscript{*}Co-first author. \quad \textsuperscript{\dag}Corresponding author.}

\maketitle              
\begin{abstract}
Alzheimer’s disease (AD) manifests as a continuous progression from normal cognition (NC) through mild cognitive impairment (MCI) to dementia. However, most deep learning approaches reduce this continuum to disjointed classification tasks, largely ignoring dynamic stage transitions. 
To decode this complex progression, we propose M\textsuperscript{3}AD, a unified framework that jointly addresses three-class diagnosis classification and diagnosis stage transition prediction using only T1-weighted sMRI. 
M\textsuperscript{3}AD leverages an interpretable multi-gate mixture of experts architecture, employing specialized routing mechanisms to dynamically capture both diagnosis-specific pathological patterns and shared structural features across the continuum. 
It further integrates clinical priors (age, sex, eTIV) via adaptive attention fusion to enhance generalization. M\textsuperscript{3}AD achieves 95.13\% accuracy, compared to 90.44\% reported by MCLNC under its original experimental setting, and 94.87\% for transition prediction. 
Crucially, analyzing the multi-gate routing reveals distinct expert activation signatures distinguishing stable from progressive MCI, providing a mechanistic basis for individual-level progression risk stratification. 
Code is available at \url{https://github.com/csyfjiang/M3AD}.
\end{abstract}

\section{Introduction}
\begin{figure}[t]
    \centering
    \includegraphics[width=0.83\linewidth]{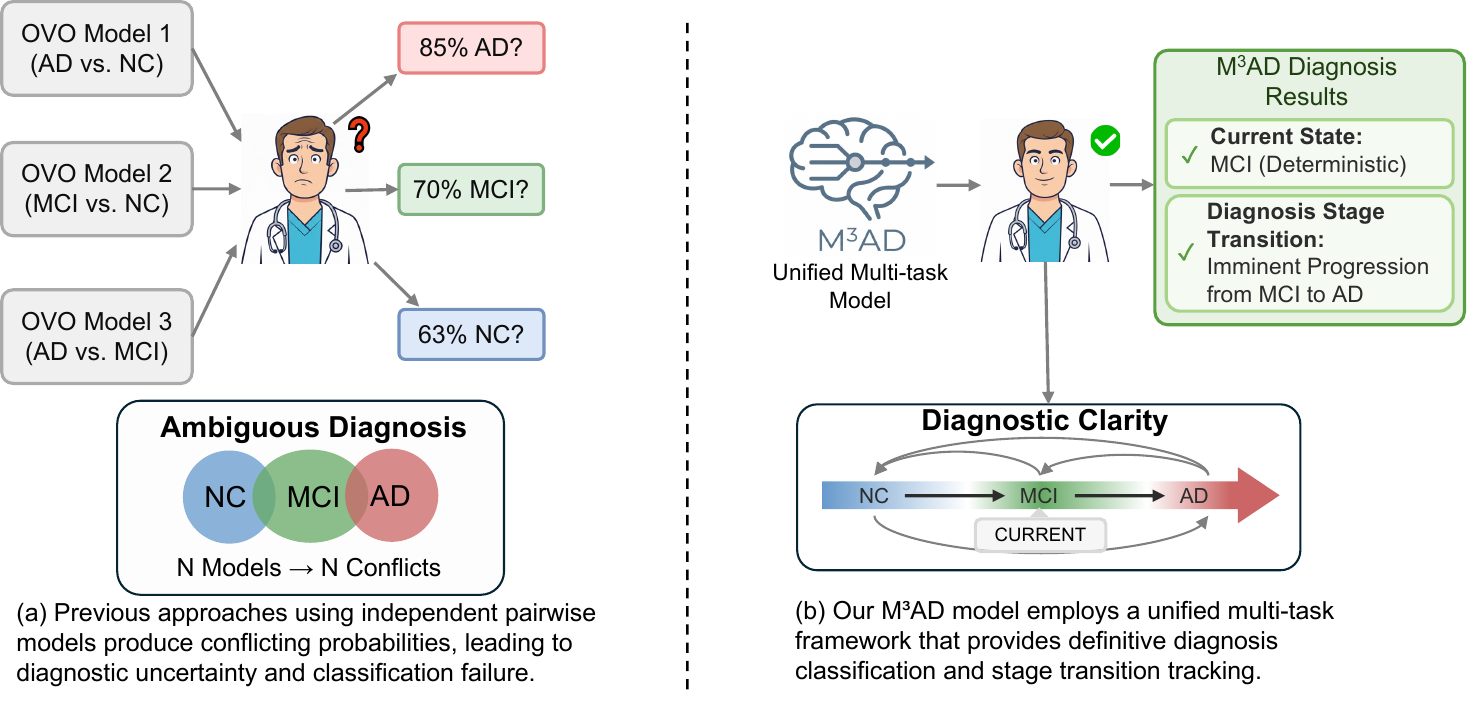}
    \caption{Comparison between the conventional one-versus-one (OVO) strategy and M\textsuperscript{3}AD. While independent OVO classifiers produce conflicting probabilities leading to diagnostic ambiguity, M\textsuperscript{3}AD resolves this via a unified multi-task model that jointly predicts the current diagnosis (NC/MCI/AD) and stage transitions along the AD continuum.}
    \label{fig:ambi}
\end{figure}

Neuro-degeneration along the Alzheimer's disease (AD) continuum, which spans from normal cognition (NC) through mild cognitive impairment (MCI) to AD dementia, manifests as quantifiable structural brain alterations~\cite{2_tahami2022alzheimer} that are effectively captured and tracked via structural MRI (sMRI)~\cite{4_hamaide2016neuroplasticity}. 
Beyond static diagnostic neuroimaging biomarkers, sMRI also detects latent indicators of stage transition patterns driven by the nature of neuroplasticity~\cite{3_chou2022cortical}. 
However, most deep learning frameworks reduce this continuum to simple two-class or three-class classification tasks~\cite{5_alorf2022multi,6_upadhyay2024comprehensive}, failing to model these clinically valuable dynamic trajectories. 
Furthermore, independent one-versus-one (OVO) classifiers~\cite{16_ding2025denseformer} often generate conflicting  probabilities due to unconstrained label overlap (Fig. \ref{fig:ambi}), fundamentally misrepresenting the mutually exclusive nature of real-world clinical assessments while remaining blind to dynamic stage transitions along the disease continuum.
To address these limitations, the multi-task Multi-gate Mixture-of-Experts (MMoE) framework leverages shared and task-specific features jointly, reducing gradient interference and improving convergence stability~\cite{9_chen2018gradnorm,10_ma2018modeling}. 
Its effectiveness in neuroimaging has been demonstrated across diverse applications~\cite{li2025m4_app,Jia_M4oE_MICCAI2024_app}. 
Moreover, to fully ground such models in clinical reality, it is crucial to incorporate clinical priors. 
Age, sex, and estimated total intracranial volume (eTIV) are established correlates of brain atrophy and AD risk, yet rarely integrated in a principled manner within existing frameworks~\cite{res18_liu2023patch}.

To address these gaps, we propose M\textsuperscript{3}AD, a \textbf{M}ulti-task \textbf{M}ulti-gate \textbf{M}ixture of experts framework built upon Swin V2~\cite{liu2021swinv2} with Tok-MLP~\cite{valanarasu2022unex_tok_mlp}. 
M\textsuperscript{3}AD jointly addresses diagnosis stage classification (NC/MCI/AD) and dynamic stage transition prediction, incorporating age, sex, and eTIV as clinical priors via adaptive attention fusion. 
Specifically, SimMIM pretraining first initializes diagnosis-specific expert specialization, followed by multi-task fine-tuning where specialized experts capture pathological patterns and shared experts model common structural features. 
Validated across six datasets comprising 12,037 sMRI scans, M\textsuperscript{3}AD achieves 95.13\% accuracy for three-class diagnosis, outperforming the state-of-the-art~\cite{zhao2024multimodal} by 4.69\% and surpassing standard vision transformer baselines including ViT (89.54\%). 
Beyond performance, interpretability analysis of the dual-gate routing mechanism reveals distinct expert activation signatures between MCI stability cases and MCI to AD progression cases, offering a mechanistic basis for individual-level progression risk stratification.

The contributions of this work are summarized as follows. 
(1) We introduce a unified M\textsuperscript{3}AD framework that jointly optimizes three-class diagnosis stage classification and diagnosis stage transition pattern prediction within a single end-to-end MMoE architecture, validated on 12,037 sMRI scans across six datasets. 
(2) We design a clinical prior integration module incorporating age, sex, and eTIV via adaptive attention fusion, enabling context-sensitive balancing of imaging and demographic information across diverse patient populations. 
(3) We present an interpretation analysis of the dual-gate routing mechanism, unveiling distinct expert activation signatures and demonstrating that our imaging–demographic fusion is consistent with established theories of functional brain networks.
\section{Methodology}

\begin{figure}[!t]
    \centering
    \includegraphics[width=0.86\linewidth]{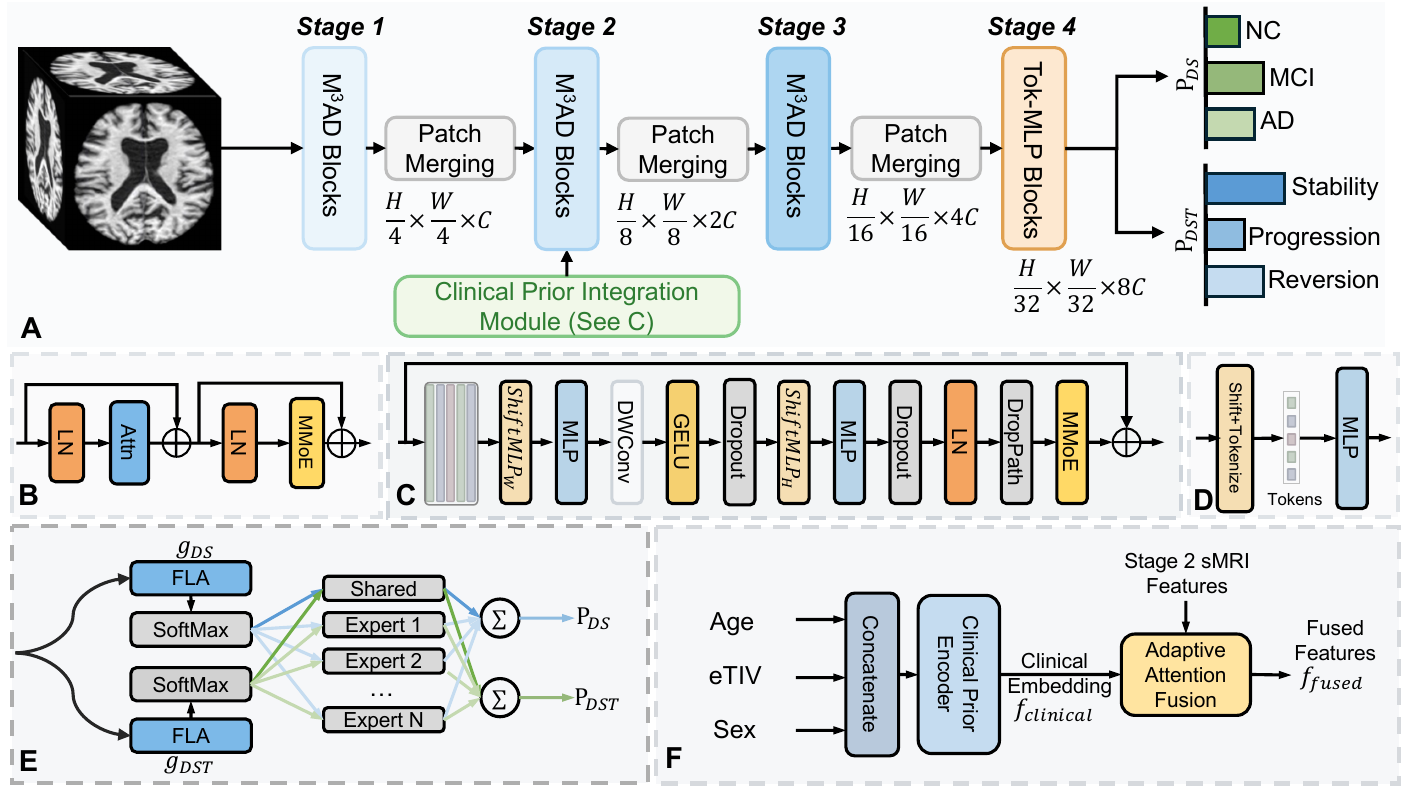}
    \caption{Overview of the M\textsuperscript{3}AD framework. (A) Four-stage hierarchical architecture processing sMRI inputs with clinical prior integration at Stage 2, producing dual outputs $P_{DS}$ (NC/MCI/AD) and $P_{DST}$ (Stability/Progression/Reversion). (B) M\textsuperscript{3}AD block with MSA and MMoE. (C) Tok-MLP block with dual ShiftMLP operations. (D) ShiftMLP module. (E) Dual gate MMoE routing via feature-level attention (FLA). (F) Clinical Prior Integration Module with adaptive attention fusion.}
    \label{fig:m3ad_architecture}
\end{figure}

\subsubsection*{Study Cohort Design}

Diagnosis stage classification used diagnosis stage labels NC, MCI, and AD. Diagnosis stage transition prediction used only the baseline first visit T1 weighted structural MRI scan and a 36 month observation window, with the endpoint defined as the last available follow up diagnosis within that window, and the diagnosis stage transition label derived from baseline and endpoint diagnoses as Stability, Progression, or Reversion.

\subsubsection*{Datasets}
This study uses six neuroimaging cohorts with 12,037 T1-weighted structural MRI scans. 
The primary cohort is ADNI\footnote{\url{https://adni.loni.usc.edu/}}, contributing 8,243 scans with diagnosis stage label set \{NC, MCI, AD\} and diagnosis stage transition labels for multi-task learning. DLBS\footnote{\url{https://fcon_1000.projects.nitrc.org/indi/retro/dlbs.html}}, IXI\footnote{\url{https://brain-development.org/ixi-dataset/}}, NKI-RS~\cite{NKI_RS}, OASIS-1~\cite{oasis1}, and OASIS-2~\cite{oasis2} contribute 3,794 additional scans to broaden acquisition variability and reduce domain shift. 
Among the six cohorts, only ADNI provides longitudinal transition annotations; the remaining five datasets contribute exclusively to the diagnosis stage classification task.
We formulate diagnosis stage transition prediction with a 3-class label set (C3: Stability, Progression, and Reversion) and diagnosis stage transition pattern prediction with a theoretically complete 9-class label set (C9) enumerating all specific inter-stage transitions. 
Due to the absence of AD Reversion cases (AD to MCI; and AD to NC) in the original dataset, C9 is empirically reduced to 7 observable classes in our implementation. 
All scans undergo standardized preprocessing with MONAI~\cite{cardoso2022monai}, N4 bias field correction, skull stripping with HD-BET~\cite{HDBET}, and SyN registration to MNI152~\cite{MNI152}. Intensity normalization uses z-score standardization within the brain foreground. All datasets were accessed under their respective data use agreements and institutional approvals.

\subsubsection*{M\texorpdfstring{\textsuperscript{3}}{3}AD Framework Overview}

Figure~\ref{fig:m3ad_architecture}A illustrates the M\textsuperscript{3}AD architecture. The model is built on a hierarchical vision transformer structure inspired by Swin V2~\cite{liu2021swinv2}. 
It processes sMRI data through four stages, where the first three stages utilize our proposed M\textsuperscript{3}AD Blocks (Figure~\ref{fig:m3ad_architecture}B). These blocks integrate an MMoE mechanism to better handle multi-task learning.
In the final stage (Stage 4), we employ Tok-MLP Blocks~\cite{valanarasu2022unex_tok_mlp} to integrate global contextual information for effective semantic mapping, facilitating the subsequent multi-task predictions. 
To enhance the model's clinical relevance, we introduce a Clinical Prior Integration Module (Figure~\ref{fig:m3ad_architecture}F). 
This module is inserted specifically at Stage 2, where it fuses clinical data (Age, Sex, eTIV) with the image features. The final output branches into two tasks: Diagnosis Stage Classification ($P_{DS}$) and Diagnosis Stage Transition Prediction ($P_{DST}$).

\subsubsection*{Base Architecture Enhanced Swin V2}

As illustrated in Fig.~\ref{fig:m3ad_architecture}B, the M\textsuperscript{3}AD block integrates the MMoE mechanism into the Swin V2~\cite{liu2021swinv2} architecture by replacing the standard feed-forward network, with the computation defined as follows:

\begin{equation}
\tilde{z}^l = \text{MSA}(\text{LN}(z^{l-1})) + z^{l-1}, \quad 
z^l_t = \text{MMoE}_t(\text{LN}(\tilde{z}^l)) + \tilde{z}^l, 
\quad t \in \{\text{DS}, \text{DST}\}.
\end{equation}
In the last stage, standard Swin V2 blocks are replaced by Tok-MLP~\cite{valanarasu2022unex_tok_mlp} as shown in Fig.~\ref{fig:m3ad_architecture}C and Fig.~\ref{fig:m3ad_architecture}D. Tok-MLP applies shifted MLP operations along spatial axes to model local atrophy patterns in sMRI
\begin{align}
&T_W = \text{Tokenize}(\text{Shift}_W(z^{l-1})), \quad \tilde{z}^l = f(\text{DWConv}(\text{MLP}(T_W))), \\
&T_H = \text{Tokenize}(\text{Shift}_H(\tilde{z}^l)), \quad z^l = f(\text{LN}(T_W + \text{MLP}(\text{GELU}(T_H)))),
\end{align}
where $\text{DWConv}$ provides positional encoding, and $\text{Shift}_W$/$\text{Shift}_H$ partition the features into $h$ groups and cyclically shift them by $j$ positions along the width and height axes respectively, inducing axial locality prior to tokenization~\cite{valanarasu2022unex_tok_mlp}.

\subsubsection*{Multi-gate Mixture of Experts (MMoE) Framework}
The MMoE uses $E=8$ experts as shown in Fig.~\ref{fig:m3ad_architecture}E. Two experts are shared across tasks, and the remaining six are organized as two diagnosis-aligned experts per class in ${\text{NC}, \text{MCI}, \text{AD}}$. Each expert maps the input by $f_e(\tilde{z}^l) = \text{MLP}_e(\text{LN}(\tilde{z}^l))$. For each task, a task-specific gate produces expert weights via feature-level attention:
\begin{equation}
g^t(\tilde{z}^l) = \text{Softmax}\!\left(\frac{\mathbf{W}_g^t \cdot \text{FLA}(\text{LN}(\tilde{z}^l))}{\tau}\right); \quad z^l_t = \sum_{e=1}^{E} g_e^t(\tilde{z}^l) \cdot f_e(\tilde{z}^l) + \tilde{z}^l,
\end{equation}
where $t \in \{\text{DS}, \text{DST}\}$ denotes diagnosis stage classification and diagnosis stage transition prediction.

\subsubsection*{Clinical Prior Integration Module}

Figure~\ref{fig:m3ad_architecture}F shows clinical priors $\mathbf{p} = [p_{\text{age}}, p_{\text{sex}}, p_{\text{eTIV}}] \in \mathbb{R}^3$. A Clinical Prior Encoder maps $\mathbf{p}$ to $\mathbf{p}_{\text{encoded}} \in \mathbb{R}^{C_{\text{fusion}}}$ and projects it to a feature tensor $\mathbf{X}_{\text{clinical}}$ compatible with Stage 2 features. Adaptive Attention Fusion computes $\mathbf{X}_{\text{fused}} = \mathbf{W}_{\text{proj}}(w_0\mathbf{X} + w_1\mathbf{X}_{\text{clinical}})$, where $(w_0, w_1) = \text{Softmax}(\text{MLP}(\text{AvgPool}([\mathbf{X}\|\mathbf{X}_{\text{clinical}}])))$.

\subsubsection*{Two-Stage Training Strategy}
\textbf{(1) SimMIM Pretraining.}
Following SimMIM~\cite{xie2022simmim}, training fold sMRI inputs are masked with ratio 0.6 and reconstructed. Pretraining is performed exclusively within each training fold, with no access to validation or test subjects, ensuring no label leakage across folds. Expert selection is guided by diagnosis stage labels, and each diagnosis-aligned expert is trained on its assigned subset. This structured initialization encourages each expert to encode distinct anatomical representations, enabling the dual gates in subsequent fine-tuning to dynamically compose shared and stage-specific knowledge, which also provides the basis for interpretable expert activation analysis
\begin{equation}
\begin{aligned}
\mathcal{L}_{\text{pretrain}} = \mathcal{L}_{\text{recon}} + \lambda \mathcal{L}_{\text{expert}}, 
\mathcal{L}_{\text{recon}} = \frac{1}{|\mathcal{M}|} \sum_{(i,j) \in \mathcal{M}} \|I_{i,j} - \hat{I}_{i,j}\|_1, 
\end{aligned}
\end{equation}
where $\mathcal{L}_{\text{expert}}$ measures the L1 reconstruction error for each diagnosis-aligned expert on its corresponding samples, encouraging stage-specific specialization.
\textbf{(2) Supervised fine-tuning.} Dual gates are enabled to dynamically route and compose the pretrained expert representations, and all parameters are optimized with a multi-task objective over diagnosis stage classification and diagnosis stage transition prediction
\begin{equation}
\mathcal{L}_{\text{finetune}} = \,\mathcal{L}_{\text{DS}} + \mathcal{L}_{\text{DST}},
\end{equation}
where $\mathcal{L}_{\text{DS}}$ is a cross entropy loss over the diagnosis stage label set \{NC, MCI, AD\} and $\mathcal{L}_{\text{DST}}$ is a cross entropy loss over the diagnosis stage transition label set $\{Stability, Progression, Reversion\}$.
\section{Experiments}

\subsubsection*{Implementation Details}
M\textsuperscript{3}AD is implemented in PyTorch 2.7.1. Training 
runs on one server with four NVIDIA H800 GPUs with 80\,GB memory 
each. Optimization uses AdamW with learning rate $10^{-4}$ for 
pretraining and $5\times10^{-5}$ for fine-tuning, cosine annealing, 
weight decay $0.05$, gradient clipping $1.0$, and mixed precision; 
the MMoE gating temperature $\tau$ is set to $1.0$ during 
pretraining and annealed to $0.5$ during fine-tuning. Full 
hyperparameter configurations are provided in the code repository. 
Training lasts $200$ epochs with batch size $256$ per GPU. We 
employ five-fold cross-validation with early stopping under 
multiple random seeds, ensuring that all longitudinal scans from 
the same subject are confined to a single fold to prevent data 
leakage. Sagittal slices are extracted and resampled to 
$256\times256$ resolution with z-score intensity normalization, 
and augmentation is restricted to neuroanatomy-preserving 
transformations. At validation, predictions from all sagittal 
slices of a subject are aggregated via majority voting to produce 
the final subject-level label. We note that slice-level labels 
constitute weak supervision derived from subject-level diagnoses; 
while majority voting mitigates the impact of unreliable individual 
slices at inference, training-time label noise from uninformative 
slices remains a limitation and will be addressed in future work.
\begin{table}[t]
\caption{Comparison with prior methods. \textbf{Left}: Three-class 
diagnosis stage classification \{NC, MCI, AD\} on pooled datasets with 
five-fold cross-validation. \textbf{Right}: Cross-cohort 
generalization evaluated as NC vs AD binary classification, trained 
on ADNI and tested on OASIS-1\&2. Results for other methods are cited from their 
original publications on respective datasets for reference. $^\dagger$ denotes SimMIM-pretrained 
initialization. Best values are shown in \textbf{bold}. ``$-$'' indicates results not reported.}
\label{tab:main}
\begin{minipage}[t]{0.48\textwidth}
\centering
\resizebox{\textwidth}{!}{%
\begin{tabular}{l|ccccc}
\hline
\textbf{Method} & Acc & Rec & Pre & Spe & F1 \\ \hline
ResNet-50~\cite{he2015deep} & 89.21 & 88.45 & 87.93 & 92.13 & 88.18 \\
DenseNet~\cite{huang2017densely} & 88.67 & 87.32 & 88.01 & 91.45 & 87.66 \\
ViT~\cite{dosovitskiy2020image} & 89.54 & 88.91 & 89.12 & 92.67 & 89.01 \\
MCAD~\cite{res6_zhang2023multi} & 64.03 & 63.85 & - & 82.00 & 61.85 \\
Stacked DAE~\cite{res17_venugopalan2021multimodal} & 78.00 & 78.00 & 77.00 & - & 78.00 \\
PDMML~\cite{res18_liu2023patch} & 80.80 & 81.00 & 81.00 & - & 81.00 \\
MCLNC~\cite{zhao2024multimodal} & 90.44 & 86.29 & 88.97 & 93.47 & 87.47 \\
\hline
MT-M$^3$AD-C3$^\dagger$ & \textbf{95.13} & \textbf{94.84} & 94.15 & \textbf{97.54} & \textbf{94.48} \\
MT-M$^3$AD-C9$^\dagger$ & 94.72 & 93.82 & \textbf{95.23} & 97.03 & 94.47 \\
\hline
\end{tabular}%
}
\end{minipage}
\hfill
\begin{minipage}[t]{0.48\textwidth}
\centering
\resizebox{\textwidth}{!}{%
\begin{tabular}{l|ccccc}
\hline
\textbf{Method} & Acc & Rec & Pre & Spe & F1 \\ \hline
ResNet-50~\cite{he2015deep} & 91.34 & 89.76 & 90.23 & 92.18 & 89.99 \\
DenseNet~\cite{huang2017densely}  & 92.47 & 91.83 & 91.56 & 93.21 & 91.69 \\
ViT~\cite{dosovitskiy2020image}       & 93.12 & 92.45 & 92.87 & 93.78 & 92.66 \\
LA-GMF \cite{2_tahami2022alzheimer}& 93.02 & - & - & - & 91.00 \\
MC-CL \cite{res3_li20223}& 93.16 & 95.00 & 94.44 & 94.44 & 94.72 \\
SSH\&LSH \cite{res5_pan2019multiscale}& 93.65 & 91.22 & - & 96.25 & - \\
MCLNC~\cite{zhao2024multimodal} & \underline{98.60} & \textbf{98.86} & \textbf{97.82} & 98.86 & \textbf{98.29} \\
\hline
MT-M$^3$AD-C3$^\dagger$ & \textbf{98.75} & 96.18 & 95.82 & \textbf{99.47} & 95.99 \\
MT-M$^3$AD-C9$^\dagger$ & 98.08 & 97.21 & 95.34 & 99.38 & 96.27 \\
\hline
\end{tabular}%
}
\end{minipage}
\end{table}

\subsubsection*{Results}
Among the baselines, ResNet-50~\cite{he2015deep}, DenseNet~\cite{huang2017densely}, and ViT~\cite{dosovitskiy2020image} are re-implemented under our experimental protocol, while MCAD~\cite{res6_zhang2023multi}, Stacked DAE~\cite{res17_venugopalan2021multimodal}, PDMML~\cite{res18_liu2023patch}, and MCLNC~\cite{zhao2024multimodal} are cited from their respective original publications for reference.
As reported in Table~\ref{tab:main} on the left side, ResNet-50 (89.21\%), DenseNet (88.67\%), and ViT (89.54\%) achieve accuracies below 90\% on diagnosis stage classification, which suggests limited sensitivity to the subtle structural differences separating NC, MCI, and AD.
Among prior specialized methods, MCAD and Stacked DAE remain weaker despite multi-modal inputs.
MCAD applies cross-modal attention to unaligned heterogeneous features and reduces volumetric representations from $128^3$ to $4^3$ through pooling, which removes localized atrophy cues required for diagnosis stage discrimination.
Stacked DAE concatenates reconstruction-oriented autoencoder features across modalities, yielding representational mismatch that degrades diagnosis stage label separability.
MCLNC reaches 90.44\% accuracy, the highest among prior methods, but its instance-level contrastive objective suppresses category-level commonalities within \{NC, MCI, AD\} and complicates optimization.
MT-M\textsuperscript{3}AD-C3 achieves 95.13\% accuracy and 94.48\% F1, improving over MCLNC by 4.69\% and exceeding the best general backbone by 5.59\%, while the specificity of 97.54\% reduces false positive diagnoses critical for screening.
MT-M\textsuperscript{3}AD-C9 attains 94.72\% accuracy with 95.23\% precision, showing that the gain persists under a nine-class diagnosis stage transition label set.
On the cross-cohort binary task (Table~\ref{tab:main} on the right side), general backbones trained on ADNI transfer poorly to OASIS, with ViT reaching only 93.12\% despite the simplified NC vs.\ AD protocol, confirming that backbone capacity alone does not ensure domain generalization.
Prior specialized methods improve upon this, yet MCLNC, while achieving the highest recall (98.86\%) and F1 (98.29\%) among competitors, underlines that subject-level contrastive objectives remain sensitive to cohort shift.
MT-M\textsuperscript{3}AD-C3 achieves 98.75\% accuracy and 99.47\% specificity, surpassing all baselines on both metrics and demonstrating that the learned representations generalize across acquisition protocols without domain adaptation.

\begin{figure}[t]
    \centering
    \includegraphics[width=1\linewidth]{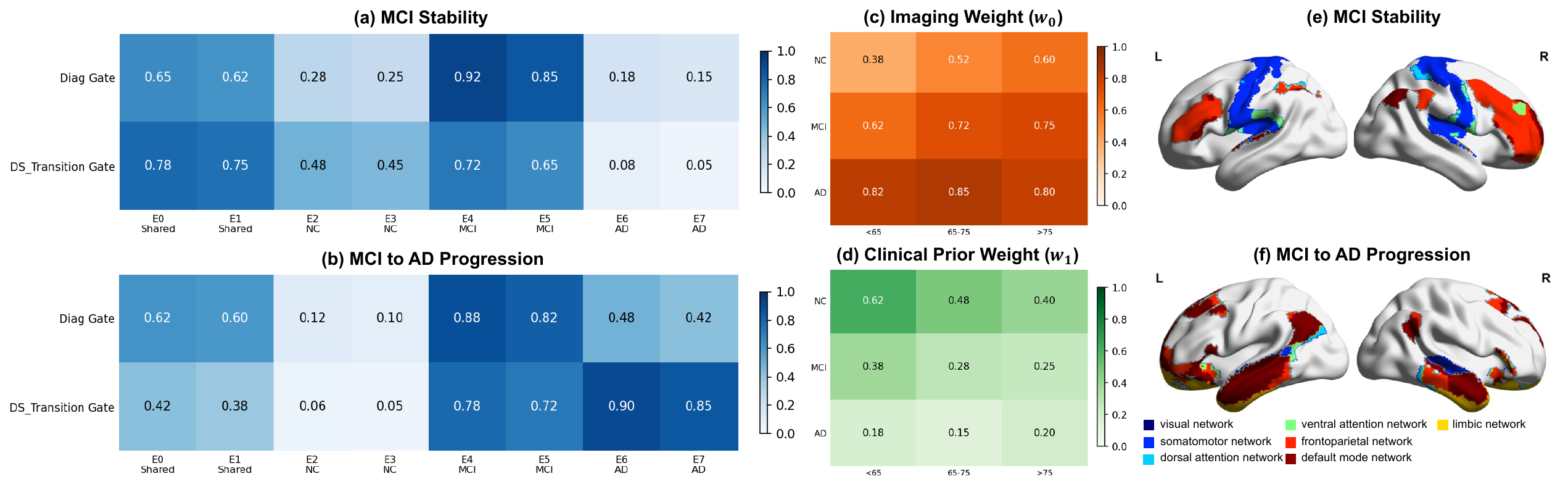}
    \caption{Interpretation analysis of M\textsuperscript{3}AD. 
     (a)(b) Expert-gate activation heatmaps contrasting the MCI Stability subgroup and the MCI to AD progression subgroup, indicating increased activation of the AD aligned expert under the diagnosis stage (DS) transition gate in the progression subgroup. 
     (c)(d) Adaptive fusion weights across diagnosis stage groups and age cohorts, where the imaging weight $w_0$ increases from NC to AD and the clinical prior weight $w_1$ decreases accordingly. 
     (e)(f) GradCAM~\cite{selvaraju2020grad} visualizations of discriminative brain regions for the MCI Stability subgroup and the MCI to AD progression subgroup, mapped into seven functional brain networks.}
    \label{fig:interpret}
\end{figure}
    
\begin{table}[!b]
\caption{Performance of M\textsuperscript{3}AD under single-task, multi-task, and ablation settings for diagnosis stage classification and transition pattern prediction on ADNI dataset. C3 and C9 denote ternary and nine-class transition annotations, respectively. Results are reported as mean$\pm$std across cross-validation folds. $^\dagger$ denotes SimMIM-pretrained initialization.}

\label{tab:m3ad_results}
\begin{minipage}[t]{0.47\textwidth}
\centering
\resizebox{\textwidth}{!}{%
\begin{tabular}{l|ll|ll}
\hline
\textbf{Method} & \multicolumn{2}{c|}{\textbf{Diagnosis}} & \multicolumn{2}{c}{\textbf{Transition Pattern}} \\ \hline
                & Acc($\pm$std)      & F1($\pm$std)      & Acc($\pm$std)        & F1($\pm$std)        \\ \hline
                & \multicolumn{4}{c}{Single Tasks}                  \\ \hline
M\textsuperscript{3}AD-C9          & 89.12(1.8)   & 88.12(1.7)   & 89.34(1.8)   & 69.21(2.8) \\
M\textsuperscript{3}AD-C3          & 89.12(1.6)   & 88.12(1.6)   & 91.23(1.6)   & 82.47(2.3) \\
M\textsuperscript{3}AD-C9$^\dagger$ & 94.80(0.8)  & 94.49(0.7)   & 91.56(1.2)   & 73.18(2.1) \\
M\textsuperscript{3}AD-C3$^\dagger$ & 94.80(0.7)  & 94.49(0.6)   & 93.45(1.0)   & 85.34(1.6) \\ \hline
                & \multicolumn{4}{c}{Multiple Tasks}                \\ \hline
M\textsuperscript{3}AD-C9          & 89.23(1.9)   & 89.25(1.8)   & 90.67(1.7)   & 71.83(2.6) \\
M\textsuperscript{3}AD-C3          & 90.15(1.7)   & 89.24(1.7)   & 92.34(1.5)   & 84.12(2.1) \\
M\textsuperscript{3}AD-C9$^\dagger$ & 94.72(0.9)  & 94.47(0.8)   & 93.21(1.0)   & 75.47(1.8) \\
M\textsuperscript{3}AD-C3$^\dagger$ & 95.13(0.8)  & 94.48(0.7)   & 94.87(0.8)   & 87.63(1.4) \\ 
w/o Tok-MLP (C3)$^\dagger$        & 93.47(1.0)  & 93.12(0.9)   & 93.56(1.1)   & 85.21(1.7) \\ \hline
\end{tabular}%
}
\end{minipage}
\hfill
\begin{minipage}[t]{0.47\textwidth}
\centering
\resizebox{\textwidth}{!}{%
\begin{tabular}{l|ll|ll}
\hline
\textbf{Method} & \multicolumn{2}{c|}{\textbf{Diagnosis}} & \multicolumn{2}{c}{\textbf{Transition Pattern}} \\ \hline
                & Acc($\pm$std)      & F1($\pm$std)      & Acc($\pm$std)        & F1($\pm$std)        \\ \hline
                & \multicolumn{4}{l}{\textit{w/o Clinical Prior}} \\ 
M\textsuperscript{3}AD-C9$^\dagger$ & 92.45(1.2) & 92.42(1.1) & 91.34(1.3) & 71.28(2.2) \\
M\textsuperscript{3}AD-C3$^\dagger$ & 93.21(1.0) & 92.72(0.9) & 92.56(1.1) & 83.47(1.8) \\ \hline
                & \multicolumn{4}{l}{\textit{w/ Single Clinical Prior (M\textsuperscript{3}AD-C3)}} \\ 
w/ Age$^\dagger$          & 94.21(0.9) & 93.87(0.8) & 93.78(0.9) & 85.34(1.5) \\
w/ Sex$^\dagger$          & 93.56(1.0) & 93.14(0.9) & 93.12(1.0) & 84.67(1.6) \\
w/ eTIV$^\dagger$         & 94.45(0.8) & 94.02(0.8) & 94.01(0.9) & 85.89(1.4) \\ \hline
                & \multicolumn{4}{l}{\textit{Different Fusion Stages (M\textsuperscript{3}AD-C3)}} \\ 
Stage 0$^\dagger$         & 93.87(1.1) & 93.28(1.0) & 93.34(1.1) & 84.56(1.7) \\
Stage 1$^\dagger$         & 94.23(0.9) & 93.82(0.8) & 94.12(0.9) & 85.78(1.5) \\
Stage 3$^\dagger$         & 94.01(1.0) & 93.60(0.9) & 93.67(1.0) & 85.12(1.6) \\ \hline
\end{tabular}%
}
\end{minipage}
\begin{flushleft}
\end{flushleft}
\end{table}

\subsubsection*{Ablation Study}
As shown in Table~\ref{tab:m3ad_results}, multi-task learning consistently outperforms single-task models. For MT-M\textsuperscript{3}AD-C3, pretraining improves diagnosis accuracy to 95.13\% and transition accuracy to 94.87\%, while removing the Tok-MLP reduces these to 93.47\% and 93.56\%, respectively. Integrating clinical priors at Stage 2 achieves optimal diagnosis accuracy, surpassing Stage 0 (93.87\%), Stage 1 (94.23\%), and Stage 3 (94.01\%). Excluding all priors drops diagnosis accuracy by 1.92\%. Among single priors, eTIV (94.45\%) provides the largest gain, followed by age (94.21\%) and sex (93.56\%). Notably, as most transition patterns concentrate on stability, a long-tail issue persists; future work will focus on addressing this class imbalance.

\subsubsection*{Interpretation Analysis}
As shown in Fig.~\ref{fig:interpret}(a)(b), the diagnosis stage transition gate in the MCI to AD progression subgroup increases activation of the AD aligned experts E6 to E7 from 0.08 to 0.90 relative to the MCI Stability subgroup. 
This gating shift indicates cross stage feature sharing that supports multi-task training and yields an individual-level progression risk signal. Fig.~\ref{fig:interpret}(c)(d) shows adaptive fusion weights across diagnosis stage groups and age cohorts. 
The imaging weight $w_0$ increases with disease severity, with NC in the range 0.38 to 0.60 and AD in the range 0.82 to 0.85, while the clinical prior weight $w_1$ is larger in younger NC subjects at 0.62 where structural differences are limited. 
Fig.~\ref{fig:interpret}(e) shows discriminative regions for the MCI Stability subgroup concentrated in the frontoparietal network (FPN) and somatomotor network (SMN). 
Fig.~\ref{fig:interpret}(f) shows discriminative regions for the MCI to AD progression subgroup involving the default mode network (DMN), limbic network (LMN), and FPN.

\section{CONCLUSION}
In this work, we present M\textsuperscript{3}AD, a unified multi-task multi-gate mixture of experts framework that jointly optimizes diagnosis stage classification \{NC, MCI, AD\} and stage transition prediction from structural MRI, integrating clinical priors via adaptive attention fusion. M\textsuperscript{3}AD achieves 95.13\% diagnosis accuracy and 94.87\% transition prediction accuracy on 12,037 scans, while its dual-gate routing mechanism reveals interpretable expert activation signatures for individualized progression risk estimation. Future work will extend cross-cohort validation and leverage routing patterns to identify structural markers of early stage transitions.

\section*{ACKNOWLEDGEMENTS}

This work was fully supported by a Collaborative Research Fund (Project No. C5082-25E), a General Research Fund (Project No. 15101422) from the Research Grants Council of the Hong Kong SAR, P. R. China, an internal grant (P0051278) of The Hong Kong Polytechnic University, and the Theme-based Research Scheme (Project No. T45-401/22-N) from the Research Grants Council of the Hong Kong SAR, P. R. China. Zongxi Li has been supported by Lingnan University through Lam Woo Research Fund (LWP20040), Direct Grant (DR26E9), and Interdisciplinary and Strategic Research Grant (ISRG252601).

\subsubsection{Disclosure of Interests} The authors have no competing interests to declare that are relevant to the content of this article. 

\bibliographystyle{splncs04}
\bibliography{ref}

@article{2_tahami2022alzheimer,
  title={Alzheimer’s disease: epidemiology and clinical progression},
  author={Tahami Monfared, Amir Abbas and others},
  journal={Neurology and therapy},
  volume={11},
  number={2},
  pages={553--569},
  year={2022},
  publisher={Springer}
}

@article{3_chou2022cortical,
  title={Cortical excitability and plasticity in Alzheimer’s disease and mild cognitive impairment: A systematic review and meta-analysis of transcranial magnetic stimulation studies},
  author={Chou, Ying-hui and Sundman, Mark and others},
  journal={Ageing research reviews},
  volume={79},
  pages={101660},
  year={2022},
  publisher={Elsevier}
}

@article{4_hamaide2016neuroplasticity,
  title={Neuroplasticity and MRI: a perfect match},
  author={Hamaide, Julie and De Groof, Geert and Van der Linden, Annemie},
  journal={NeuroImage},
  volume={131},
  pages={13--28},
  year={2016},
  publisher={Elsevier}
}

@article{5_alorf2022multi,
  title={Multi-label classification of Alzheimer's disease stages from resting-state fMRI-based correlation connectivity data and deep learning},
  author={Alorf, Abdulaziz and Khan, Muhammad Usman Ghani},
  journal={Computers in Biology and Medicine},
  volume={151},
  pages={106240},
  year={2022},
  publisher={Elsevier}
}

@article{6_upadhyay2024comprehensive,
  title={Comprehensive systematic computation on Alzheimer's disease classification},
  author={Upadhyay, Prashant and Tomar, Pradeep and Yadav, Satya Prakash},
  journal={Archives of Computational Methods in Engineering},
  volume={31},
  number={8},
  pages={4773--4804},
  year={2024},
  publisher={Springer}
}

@inproceedings{9_chen2018gradnorm,
  title={Gradnorm: Gradient normalization for adaptive loss balancing in deep multitask networks},
  author={Chen, Zhao and Badrinarayanan, Vijay and Lee, Chen-Yu and Rabinovich, Andrew},
  booktitle={International conference on machine learning},
  pages={794--803},
  year={2018},
  organization={PMLR}
}

@inproceedings{10_ma2018modeling,
  title={Modeling task relationships in multi-task learning with multi-gate mixture-of-experts},
  author={Ma, Jiaqi and Zhao, Zhe and others},
  booktitle={Proceedings of the 24th ACM SIGKDD international conference on knowledge discovery \& data mining},
  pages={1930--1939},
  year={2018}
}

@article{16_ding2025denseformer,
  title={DenseFormer-MoE: A Dense Transformer Foundation Model with Mixture of Experts for Multi-Task Brain Image Analysis},
  author={Ding, Rizhi and Lu, Hui and Liu, Manhua},
  journal={IEEE TMI},
  year={2025},
  publisher={IEEE}
}

@InProceedings{Jia_M4oE_MICCAI2024_app,
        author = { Jiang, Yufeng and Shen, Yiqing},
        title = { { M4oE: A Foundation Model for Medical Multimodal Image Segmentation with Mixture of Experts } },
        booktitle = {proceedings of Medical Image Computing and Computer Assisted Intervention -- MICCAI 2024},
        year = {2024},
        publisher = {Springer Nature Switzerland},
        volume = {LNCS 15012},
        month = {October},
        page = {621 -- 631}
}

@article{li2025m4_app,
  title={M4: Multi-proxy multi-gate mixture of experts network for multiple instance learning in histopathology image analysis},
  author={Li, Junyu and Zhang, Ye and Shu, Wen and Feng, Xiaobing and Wang, Yingchun and Yan, Pengju and Li, Xiaolin and Sha, Chulin and He, Min},
  journal={Medical Image Analysis},
  volume={103},
  pages={103561},
  year={2025},
  publisher={Elsevier}
}

@inproceedings{xie2022simmim,
  title={Simmim: A simple framework for masked image modeling},
  author={Xie, Zhenda and Zhang, Zheng and others},
  booktitle={Proceedings of the IEEE/CVF conference on CVPR},
  pages={9653--9663},
  year={2022}
}

@inproceedings{liu2021swinv2,
  title={Swin transformer v2: Scaling up capacity and resolution},
  author={Liu, Ze and Hu, Han and others},
  booktitle={Proceedings of the IEEE/CVF conference on CVPR},
  pages={11999--12009},
  year={2022}
}

@inproceedings{valanarasu2022unex_tok_mlp,
  title={Unext: Mlp-based rapid medical image segmentation network},
  author={Valanarasu, Jeya Maria Jose and Patel, Vishal M},
  booktitle={International conference on MICCAI},
  pages={23--33},
  year={2022},
  organization={Springer}
}

@article{oasis1,
  author    = {Marcus, Daniel S. and others},
  title     = {Open Access Series of Imaging Studies ({OASIS}): Cross-sectional {MRI} Data in Young, Middle Aged, Nondemented, and Demented Older Adults},
  journal   = {Journal of Cognitive Neuroscience},
  year      = {2007},
  volume    = {19},
  number    = {9},
  pages     = {1498--1507},
}

@article{oasis2,
  title={Open access series of imaging studies: longitudinal MRI data in nondemented and demented older adults},
  author={Marcus, Daniel S and Fotenos, Anthony F and Csernansky, John G and Morris, John C and Buckner, Randy L},
  journal={Journal of cognitive neuroscience},
  volume={22},
  number={12},
  pages={2677--2684},
  year={2010},
  publisher={MIT Press One Rogers Street, Cambridge, MA 02142-1209, USA journals-info~…}
}

@article{NKI_RS,
  title={A longitudinal resource for studying connectome development and its psychiatric associations during childhood},
  author={Tobe, Russell H and MacKay-Brandt, Anna and others},
  journal={Scientific data},
  volume={9},
  number={1},
  pages={300},
  year={2022},
  publisher={Nature Publishing Group UK London}
}

@article{cardoso2022monai,
  title={Monai: An open-source framework for deep learning in healthcare},
  author={Cardoso, M Jorge and Li, Wenqi and others},
  journal={arXiv preprint arXiv:2211.02701},
  year={2022}
}

@article{MNI152,
    title = {Unbiased average age-appropriate atlases for pediatric studies},
    author = {Fonov, Vladimir and Evans, Alan C and others},
    journal = {NeuroImage},
    volume = {54},
    number = {1},
    pages = {313--327},
    year = {2011}
}

@article{HDBET,
  title={Automated brain extraction of multisequence MRI using artificial neural networks},
  author={Isensee, Fabian and Schell, Marianne and others},
  journal={Human brain mapping},
  volume={40},
  number={17},
  pages={4952--4964},
  year={2019},
  publisher={Wiley Online Library}
}

@article{res3_li20223,
  title={3-D CNN-based multichannel contrastive learning for Alzheimer’s disease automatic diagnosis},
  author={Li, Jiaguang and Wei, Ying and Wang, Chuyuan and Hu, Qian and Liu, Yue and Xu, Long},
  journal={IEEE Transactions on Instrumentation and Measurement},
  volume={71},
  pages={1--11},
  year={2022},
  publisher={IEEE}
}

@article{res5_pan2019multiscale,
  title={Multiscale spatial gradient features for 18F-FDG PET image-guided diagnosis of Alzheimer’s disease},
  author={Pan, Xiaoxi and Alzheimer’s Disease Neuroimaging Initiative and others},
  journal={Computer Methods and Programs in Biomedicine},
  volume={180},
  pages={105027},
  year={2019},
  publisher={Elsevier}
}

@article{res6_zhang2023multi,
  title={Multi-modal cross-attention network for Alzheimer’s disease diagnosis with multi-modality data},
  author={Zhang, Jin and He, Xiaohai and others},
  journal={Computers in biology and medicine},
  volume={162},
  pages={107050},
  year={2023},
  publisher={Elsevier}
}

@inproceedings{zhao2024multimodal,
  title={Multimodal contrastive learning with neuroimaging and cognitive tests for Alzheimer’s disease diagnosis},
  author={Zhao, Liang and Zhang, Jian and others},
  booktitle={2024 IEEE International Conference on BIBM},
  pages={2971--2976},
  year={2024},
  organization={IEEE}
}

@article{res17_venugopalan2021multimodal,
  title={Multimodal deep learning models for early detection of Alzheimer’s disease stage},
  author={Venugopalan, Janani and Tong, Li and others},
  journal={Scientific reports},
  volume={11},
  number={1},
  pages={3254},
  year={2021},
  publisher={Nature Publishing Group UK London}
}

@article{res18_liu2023patch,
  title={Patch-based deep multi-modal learning framework for Alzheimer’s disease diagnosis using multi-view neuroimaging},
  author={Liu, Fangyu and Yuan, Shizhong and Li, Weimin and Xu, Qun and Sheng, Bin},
  journal={Biomedical Signal Processing and Control},
  volume={80},
  pages={104400},
  year={2023},
  publisher={Elsevier}
}

@article{dosovitskiy2020image,
  title   = {An Image is Worth 16x16 Words: Transformers for Image Recognition at Scale},
  author  = {Alexey Dosovitskiy and Lucas Beyer and others},
  year    = {2020},
  journal = {arXiv preprint arXiv: 2010.11929
        
        }
}

@inproceedings{huang2017densely,
  title={Densely connected convolutional networks},
  author={Huang, Gao and Liu, Zhuang and others},
  booktitle={Proceedings of the IEEE conference on CVPR},
  pages={4700--4708},
  year={2017}
}

@article{he2015deep,
  title     = {Deep Residual Learning for Image Recognition},
  author    = {Kaiming He and X. Zhang and Shaoqing Ren and Jian Sun},
  journal   = {CVPR},
  year      = {2016},
  doi       = {10.1109/cvpr.2016.90},
}

@article{selvaraju2020grad,
  title={Grad-CAM: visual explanations from deep networks via gradient-based localization},
  author={Selvaraju, Ramprasaath R and Cogswell, Michael and Das, Abhishek and Vedantam, Ramakrishna and Parikh, Devi and Batra, Dhruv},
  journal={International journal of computer vision},
  volume={128},
  number={2},
  pages={336--359},
  year={2020},
  publisher={Springer}
}

\end{document}